\documentclass[nofootinbib,twocolumn,superscriptaddress]{revtex4-1}
\usepackage{graphicx,epsfig}
\usepackage{graphics,dcolumn,bm,epic, eepic,fleqn,float}
\usepackage{amssymb,amsmath,amsfonts,multirow,rotate,color,xcolor}
\usepackage{soul}
\usepackage{url}

\usepackage{subfigure}

\begin{document}
\author{Chlo\"e Brown}
\affiliation{Computer Laboratory, University of Cambridge, Cambridge (UK)}

\author{Christos Efstratiou}
\affiliation{School of Engineering \& Digital Arts, University of Kent (UK)}
\altaffiliation{Data collection was undertaken while the authors were at the Computer Laboratory of the University of Cambridge.}

\author{Ilias Leontiadis}
\affiliation{Telefonica Research, Barcelona (Spain)}
\altaffiliation{Data collection was undertaken while the authors were at the Computer Laboratory of the University of Cambridge.}

\author{Daniele Quercia}
\affiliation{Yahoo Labs, Barcelona (Spain)}
\altaffiliation{Data collection was undertaken while the authors were at the Computer Laboratory of the University of Cambridge.}

\author{Cecilia Mascolo}
\affiliation{Computer Laboratory, University of Cambridge, Cambridge (UK)}

\author{James Scott}
\affiliation{Microsoft Research, Cambridge (UK)}

\author{Peter Key}
\affiliation{Microsoft Research, Cambridge (UK)}
\date{\today}

\begin{abstract}The layouts of the buildings we live in shape our everyday lives. In office environments, building spaces affect employees' communication, which is crucial for productivity and innovation. However, accurate measurement of how spatial layouts affect interactions is a major challenge and traditional techniques may not give an objective view.We measure the impact of building spaces on social interactions using wearable sensing devices. We study a single organization that moved between two different buildings, affording a unique opportunity to examine how space alone can affect interactions. The analysis is based on two large scale deployments of wireless sensing technologies: short-range, lightweight RFID tags capable of detecting face-to-face interactions. We analyze the traces to study the impact of the building change on social behavior, which represents a first example of using ubiquitous sensing technology to study how the physical design of two workplaces combines with organizational structure to shape contact patterns.\end{abstract}
\title{The architecture of innovation:\\Tracking face-to-face interactions with ubicomp technologies}
\maketitle
\section{Introduction}
In the field of architecture, the effect of the nature and layout of spaces on the behavioral patterns of people is an important factor in building design. Significant effort has been put into understanding how the physical space of workplaces can directly affect how often employees meet one another and interact face-to-face~\cite{Allen06:Organization}. Communication between employees is a vital factor in the operation of an organization, and even in today's technologically connected world, face-to-face interactions remain crucial for the exchange of ideas and information~\cite{Pentland12:New,Stryker12:Facilitating}. It is therefore unsurprising that building spaces that facilitate such interactions is a significant consideration in architectural design.

Measuring the impact of a workplace building layout on face-to-face communication is an important step, not only to validate architects' objectives, but also to enable the evaluation and reconsideration of traditional design principles. Studies in architectural design, such as the work of Thomas Allen~\cite{Allen06:Organization}, consider how organizational structure and spatial configuration of work environments combined to influence communication between employees. However, these studies suffer from a crucial shortcoming: they lack reliable means of measuring face-to-face interactions in the workplace. Traditional approaches to evaluating the use of spaces in buildings rely on ethnographic studies where observers track employees over a period of time, or on self reports and surveys. Both approaches can deliver biased results, either because participants adapt their behavior when they know they are being observed~\cite{Whyte43:Street}, or because they tend to offer socially desirable responses to surveys~\cite{Bradburn78:Question}. Furthermore, studying the impact of a building's layout on social behavior is challenging considering the large number of variables that can affect such behavior. For example, different types of organizational structure may affect social behavior more significantly than space layout. 

In this work we perform a study that addresses these two challenges. Firstly, the study utilizes wearable sensing tags capable of capturing face-to-face interactions and the actual locations of people. The tags are unobtrusive and thus allow us to capture the real behavior of employees. Secondly, the study was performed in a research institution in the UK that moved from their old premises to a new purpose-designed building. Two data collection deployments were performed, one in the old building and one in the new. Considering that the set of additional variables, such as organizational structure, remained unchanged, the results allow us to study the impact that spatial design has on social behavior.

The work relies on the theoretical premise established by Thomas Allen, but in this case the analysis is based on behavior sensed using wireless tags. Allen's foundational work defines three types of communication necessary in an organization~\cite{Allen06:Organization}. The first is \emph{communication for coordination}, which takes place between people working on the same project, in order to coordinate work activities. Second, \emph{communication for information} is necessary for people working in the same area to keep up to date with developments in their field of expertise. It is intuitive that these two kinds of communication should, in a typical office environment, take place in offices and designated meeting rooms, since the managers of most organizations tend to be aware that these types of communication are crucial. When deciding who sits, where, they tend to arrange that people working on the same projects and in related fields are near to one another.

The third type of communication is \emph{communication for inspiration}, which, ``In an organization that relies on creative solutions to problems," Allen writes, ``is absolutely critical. It is usually spontaneous and often occurs between people who work in different organizational units, on different projects." The criticality of these interactions between members of different teams, who might not normally encounter one another during their work, has been demonstrated by much other research~\cite{Burt04:Structural,Pentland12:New}. It follows that as well as offices and meeting rooms, workplaces should include informal spaces such as coffee areas, where unplanned encounters between employees can take place, outside those meetings that would be expected given the formal organizational management structure and division into subgroups working on various projects~\cite{Kraut90:Informal,Waber10:Productivity}. Indeed, this idea is already being put into practice by high-tech organizations such as Google, where ``even the length of the lines inside the cafeteria are designed to make sure Google employees talk to others they don't necessarily work with [\ldots] if there is no line, you won't talk to anyone, you won't interact"~\cite{Henn13:Serendipitous}.

By performing two large scale deployments of a face-to-face and location sensing technology we have captured a unique dataset comprising interactions and location traces of the same employees working in two different buildings. We analyze the relationship between the participants' positions in the formal organizational structure, and their interactions in the differing spaces of the two different workplace buildings, and show that \emph{there are differences in observed communication patterns given the different physical spaces available} in the two buildings. 
Our contributions are summarized below:
\begin{itemize}
\item We demonstrate the feasibility of analyzing the impact of building layouts on social interactions, using wireless sensing technologies and without the need for self-reported information by the participants.
\item We evaluate the impact of building layouts through the analysis of the social behavior of the same set of people within two different building layouts. To the best of our knowledge this is the first such study using wireless sensing technologies.
\item We validate that specific architectural design decisions to facilitate communication for inspiration, such as the use of common areas for coffee and food, have a strong impact on social interactions, potentially more than the allocation of office spaces in the working environment.
\end{itemize}

\section{Related work}
\subsection{Ubiquitous sensing of workplace interactions}
The idea that location-based sensing services have much to offer in the workplace is not a new one, with this being a key motivation behind the development of the Active Badge system in the early 1990s~\cite{Want92:ActiveBadge}. This early use of such technology in business environments was centered around the design of context-aware systems, and focused less on the use of ubiquitous sensing technology to gain insight into social interactions in the workplace.

The `sociometric badges' described by Olgu\'in-Olgu\'in \emph{et al.}~\cite{Olguin09:Sensible} served as a concrete demonstration that wearable computing devices can be used to measure face-to-face interactions in the workplace. They showed that data collected in such a way has great potential for use by organizations to improve performance, and enhance interactions between employees; in this particular study they found that, combined with information about email communication, the sensed data could be used to predict people's perceptions of group interactions. These earlier sensing devices were bigger and more noticeable than the lightweight badges we use in this work, but nevertheless have been used very effectively in many studies of workplace social interactions. For example, Waber \emph{et al.}~\cite{Waber10:Productivity} investigated how social group strength can affect employee productivity, and Wu \emph{et al.}~\cite{Wu08:Mining} studied the relationship between the electronic communication network and that formed by face-to-face contacts.

Other technology used to study face-to-face interactions in the workplace includes wearable cameras, as used in very recent work by Mark \emph{et al.}~\cite{Mark14:Capturing}, and mobile phones~\cite{Efstratiou12:Sense}. While mobile phones are can sense face-to-face interactions less accurately than wearable devices such as the badges we have used in this study, mobile applications to track social interactions have the potential to go further than facilitating the analysis of collected data; they can also feed back the sensed information to users, perhaps with the aim to change their behavior.

\subsection{How the physical workspace affects communication}
In architecture, the effect of the nature and layout of spaces on behavioral patterns and group interactions is important to consider when designing buildings. Various methods exist for the analysis of such phenomena, including \emph{space syntax}, used by Penn \emph{et al.}~\cite{Penn99:Space} to show that the physical space of the workplace itself can directly affect how often employees meet one another and interact face-to-face. A 2008 study by Toker and Gray~\cite{Toker08:Innovation} concerned specifically the kind of research environment that we study here, and similarly showed that the spatial configuration of the working environment has a strong effect on the frequency and location of informal meetings between colleagues. 

Allen and Henn have collaborated extensively to show how crucial the architecture of the technical workplace is for communication within and between teams, with physical space being a management tool as important as organizational structure for today's technical organizations~\cite{Allen06:Organization}. They have further demonstrated that the interplay between these two factors can have profound effects on the process of information flow in workplace communication networks, with consequences for innovation and productivity.

\subsection{The importance of informal encounters}
The value of contact with those one might not communicate closely with on a regular basis (`weak ties') is well-known, and explained in a sociological context by Granovetter in his seminal 1973 paper~\cite{Granovetter73:Strength}. The application of the concept to employees in a workplace environment is clear, and the development of technologies that can measure who communicates with whom during the working day has allowed these effects to be measured. Notably, Pentland \emph{et al.}~\cite{Pentland12:New} recently studied communications between workers at a Prague bank, and found that teams that communicated informally outside of their working groups showed better performance than those whose members did not. They also showed the benefit of informal communication between people on the same team, through changing the coffee break structure so that teammates took their breaks at the same time and had the opportunity to interact outside of the formal working context. The same idea motivated work by Kirkham \emph{et al.}~\cite{Kirkham13:Break}, involving the implementation of a `break-time barometer' designed to use an ambient persuasion approach to encourage colleagues working in different parts of the building to take their coffee breaks at the same time, to allow simultaneous occupancy of informal spaces and the opportunity for social encounters.
\section{The impact of building spaces on interactions}
\begin{figure}
    \subfigure[Old building]{\label{fig:oldbuilding}
    \includegraphics[width=.53\columnwidth]{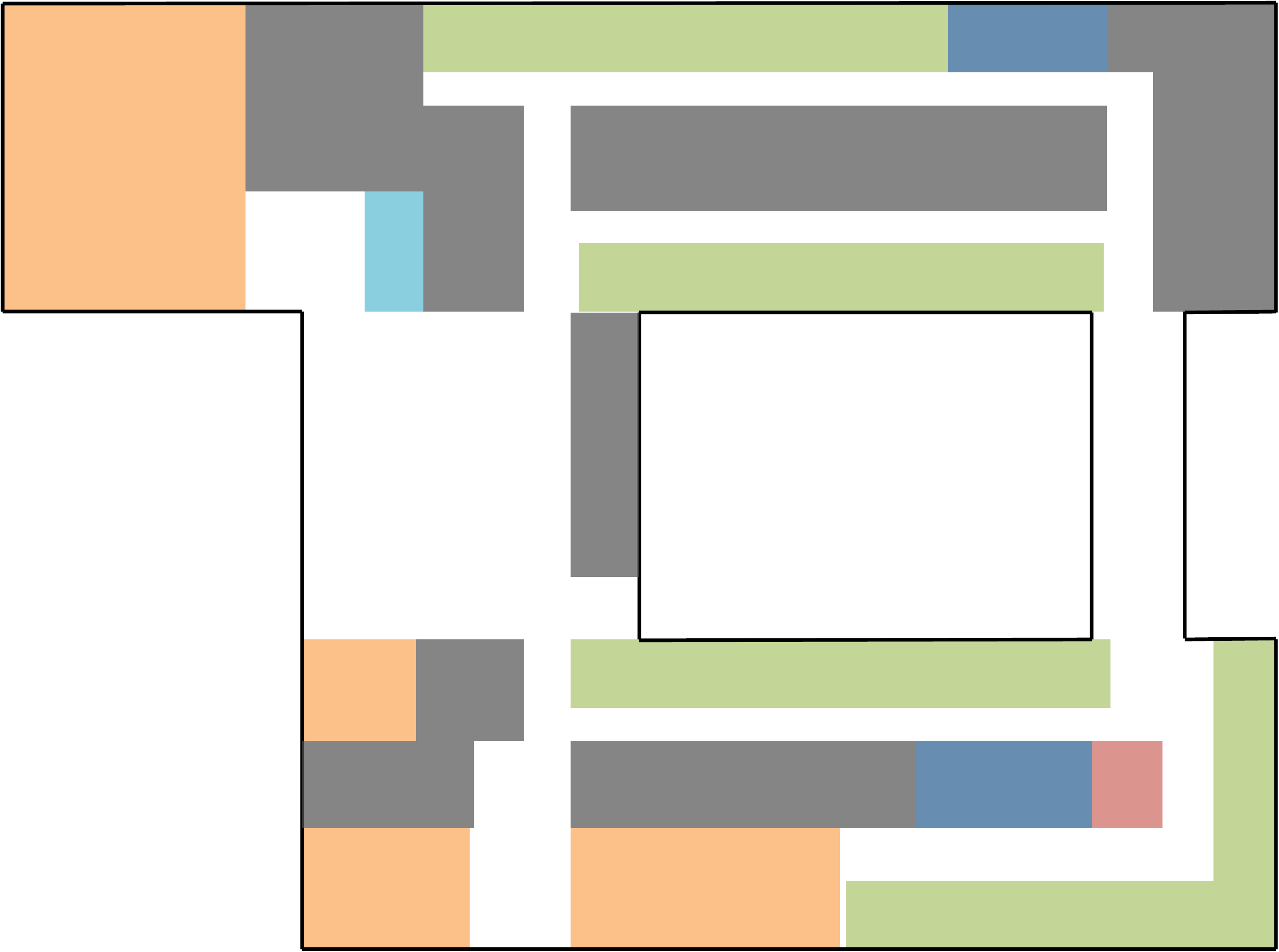}
        \hfill
        \includegraphics[width=.435\columnwidth]{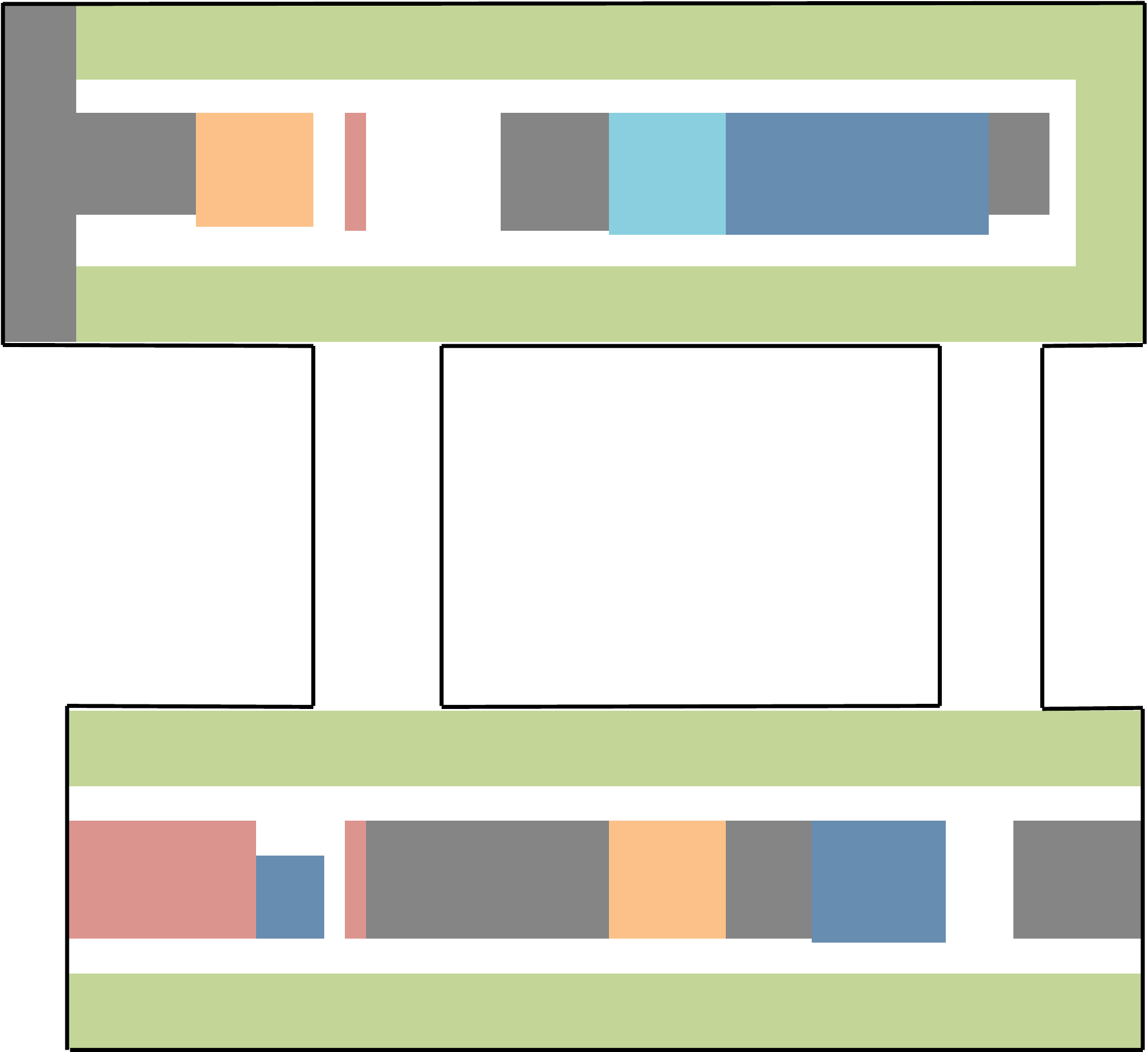}}	    
    \hfill
    \subfigure[New building]{\label{fig:newbuilding}
        \includegraphics[width=.675\columnwidth, angle=270]{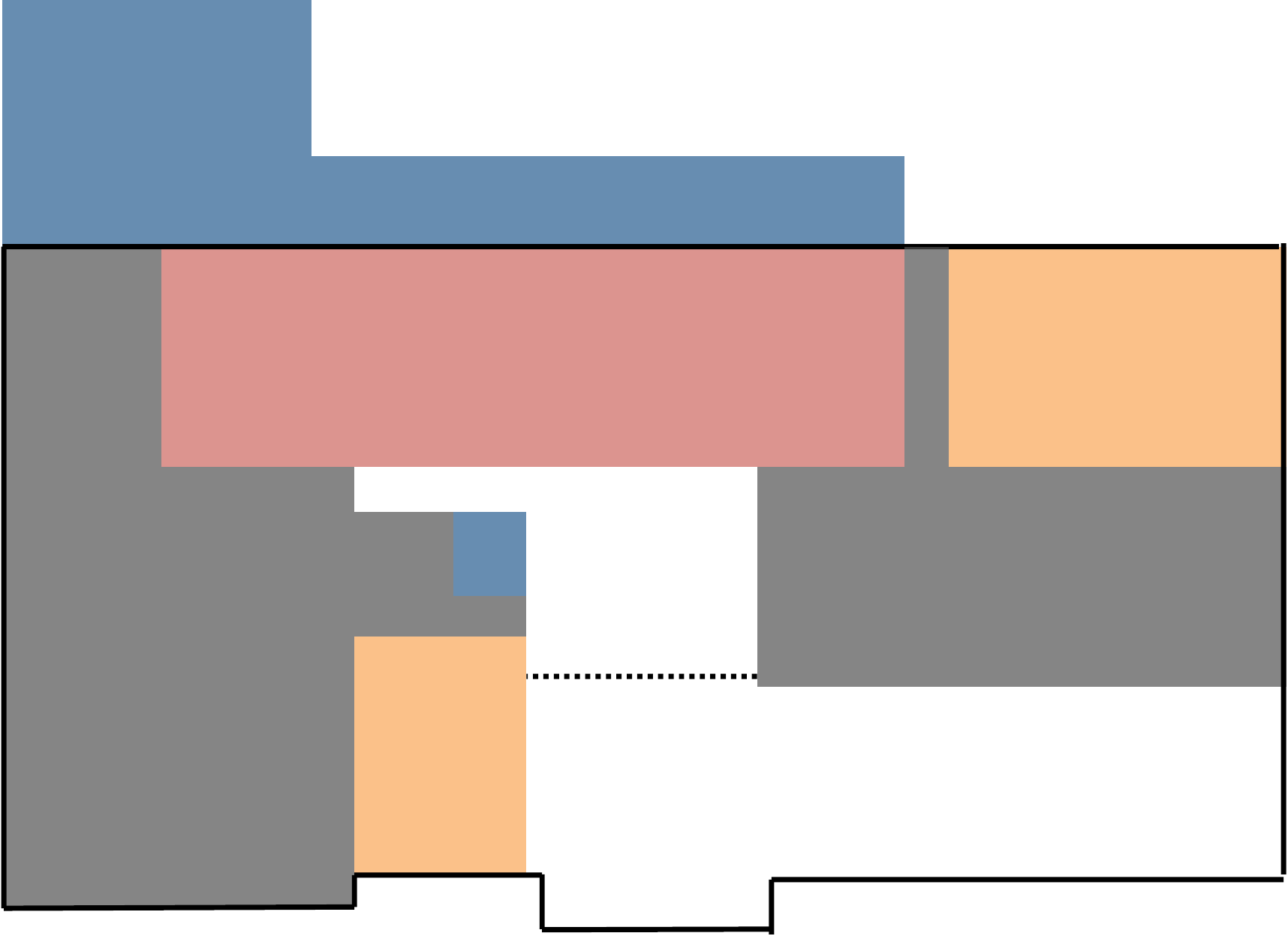}
        \hfill
        \includegraphics[width=.675\columnwidth, angle=270]{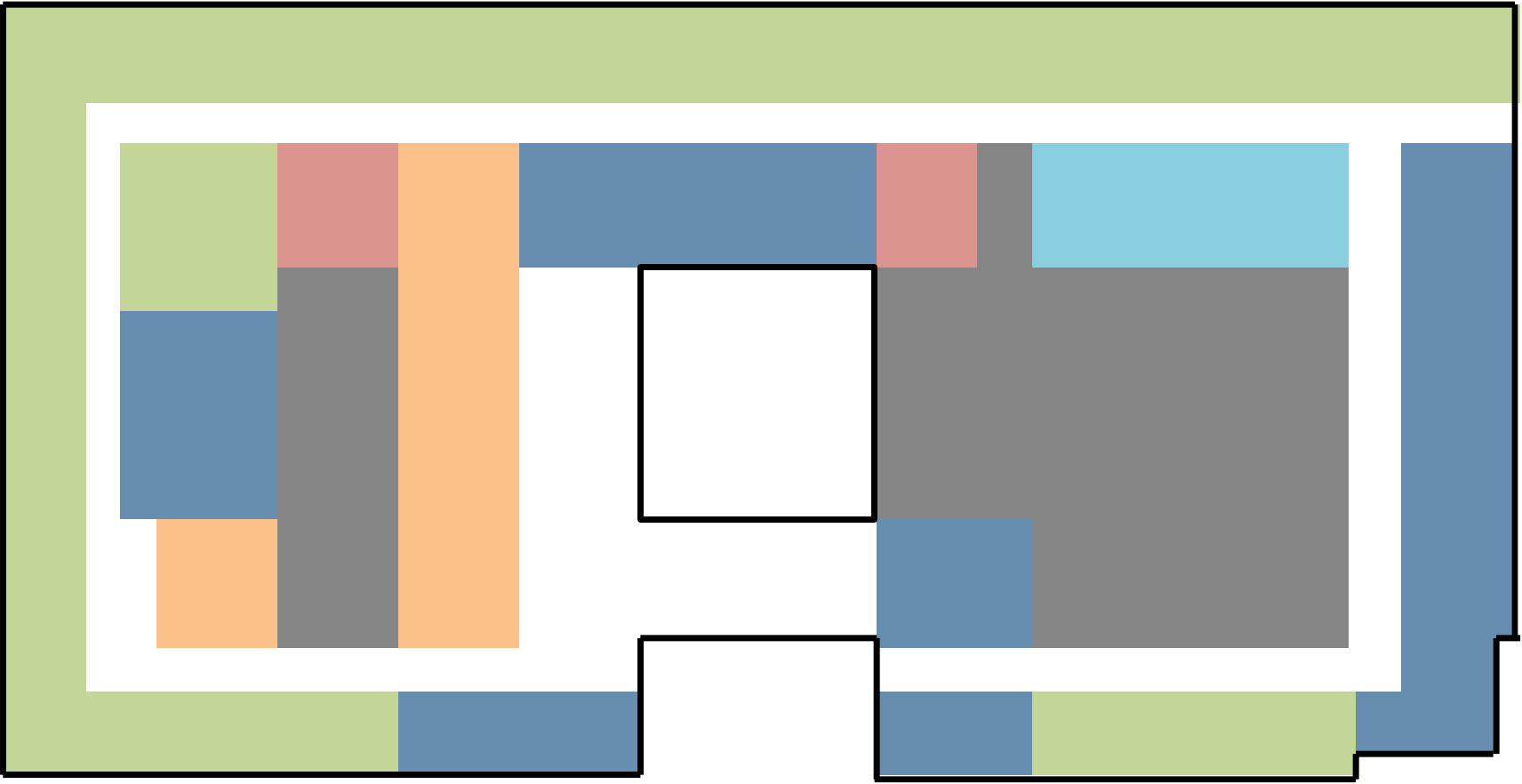}
        }
    \hfill
    \subfigure{
    \includegraphics[width=.18\columnwidth]{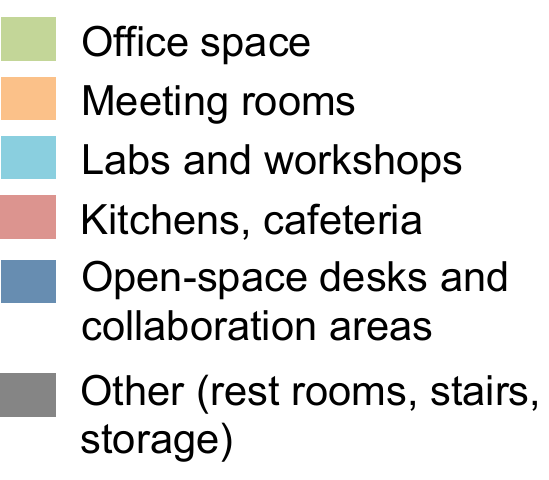}}
    \caption{Building layouts. The figures show the ground floor (left) and first floor (right) of the two buildings. Different colours indicate the type of space in each building, e.g. offices, meeting rooms, kitchens.}
    \label{fig:buildings}
\end{figure}
We conduct an empirical study of how the physical space of a workplace building can affect face-to-face interaction between employees. The study was carried out in a research institution in the UK, making use of a unique opportunity to collect suitable data, afforded by the institution moving from one building to another during the study period. This enabled us to collect data in the same way using the same technology over two periods of two weeks each: first at the old premises, and again after the organization had moved into the new building. 

\subsection{Aims in design of the new building}
The new building was designed and built specifically for the research institution in question, with the architects having particular intentions with respect to the use of the space. Through consultation with members of the design team we collected the key design decisions that focused on enabling more interaction between people from different research groups who might not usually encounter one another: in terms of Allen's types of communication, \emph{communication for inspiration}.

The most obvious difference between the two buildings in this respect is the presence in the new building of a central cafeteria area located away from the office spaces, where employees can buy food at lunchtime, and meet for coffee (see Figure~\ref{fig:buildings}).
The best two coffee machines, having coffee of the same quality as in commercial coffee houses, were deliberately placed on the ground floor, opposite to the main entrance, so that most people would have to walk past in the morning. It was expected that the quality factor would encourage people to gather in the cafeteria for good coffee, where they would have a greater chance of serendipitous encounters than in the smaller kitchens upstairs. The kitchens on the individual floors were not provided with equivalent quality machines to bring this into effect.
In the old building, there was no cafeteria serving food as in the new building; instead people would commonly buy food from elsewhere, or bring it from home, and eat it in kitchen spaces close to their offices. There was a kitchen where many people would eat, but it was not the same as having one central cafeteria as people would also eat in other spaces throughout the building. 

To a similar end, there was also a general aim in the design of the building to encourage increased use of shared spaces, as opposed to individual offices. Lab spaces were made bigger in the new building so that they might accommodate more people from different groups. There are lots of open areas and mini conference rooms without doors, in order to encourage groups to meet in these shared spaces, rather than in their own offices. It is probable that most meetings in these kinds of spaces would be related to work, and thus likely to be for the purpose of \emph{communication for information} and \emph{communication for coordination}.

In the following, we investigate whether these differences in the physical space of the new building are reflected in the patterns of interactions between its occupants. Since long-standing habits are by their nature difficult to change, behavioral differences might not be observed despite the design of the new space. We analyze traces of face-to-face contact between employees and records of their locations within the building space, in order to answer three questions regarding the effect on communication of formal organizational structure and its interaction with physical building space, as follows:

\begin{itemize}
\item \emph{To what extent is the vertical structure of the formal organizational hierarchy reflected by face-to-face interaction patterns?}
\item \emph{Is there more face-to-face communication between the different subgroups, as defined by the management hierarchy, in the new building than in the old building?}
\item \emph{Does mixing between people in different subgroups takes place in food and drink areas?}
\end{itemize}

In order to address these research questions we utilized wearable RFID tags that were able to track face-to-face interactions in the two buildings.

\section{Data collection}
We conducted an experimental study of how the physical space of a workplace can impact face-to-face interaction patterns. In particular we monitored the social behavior of the workers of a technology research institution in the United Kingdom. The company was moving from an old building to a new, purpose-built building. Taking advantage of this opportunity, we conducted two field studies where we collected data about the workers' social behavior, once in the old building, and a second one after the organization had moved. The studies were approved by the ethics committee of the University of Cambridge, and the process and consent forms were vetted by the legal representative and the privacy manager of the organization.

\begin{figure}
	\centering
	\includegraphics[width=.33\columnwidth]{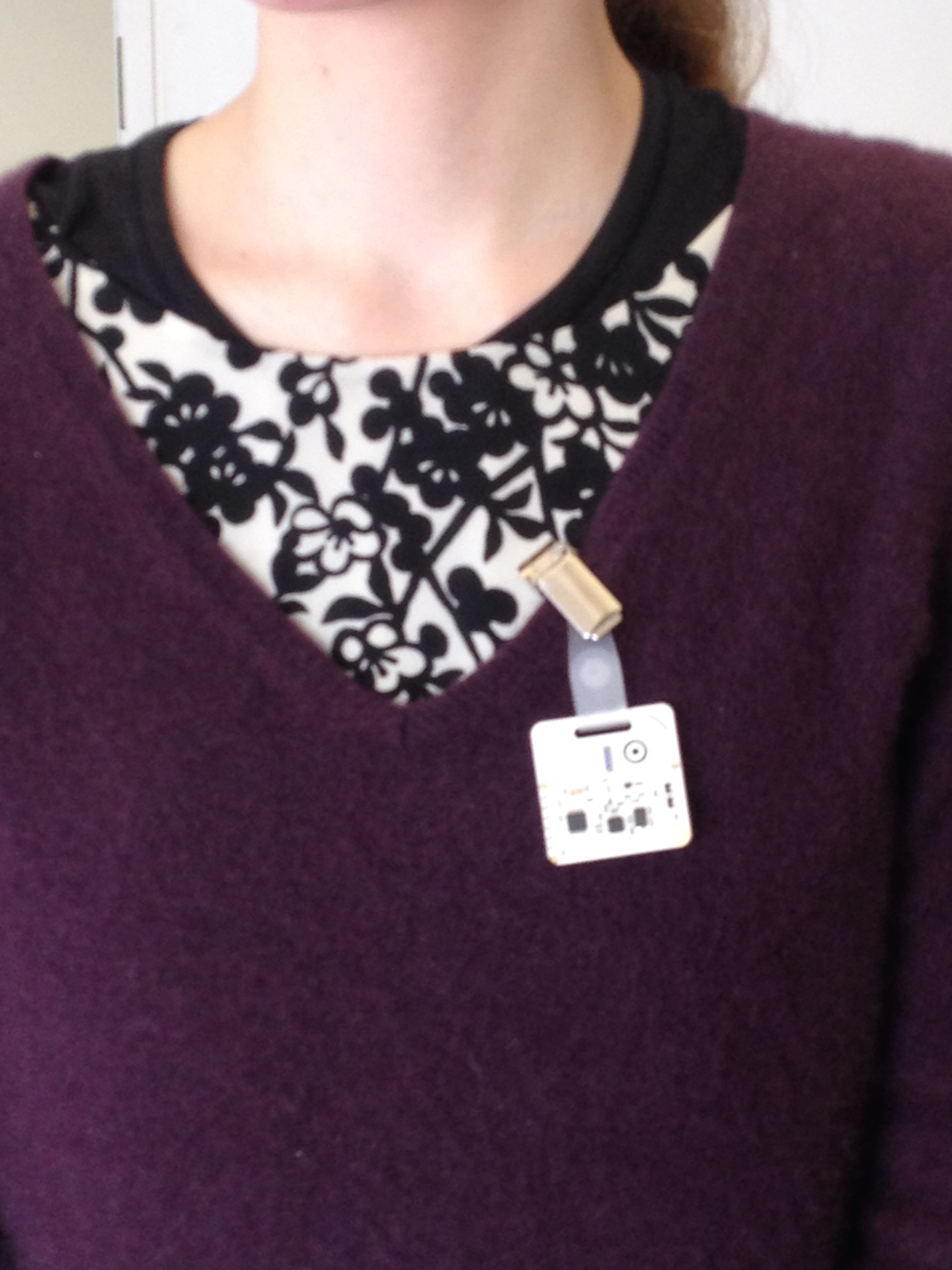}
	\caption{The RFID tags were worn on the chest of the participants, enabling communication with other tags when two participants were facing each other.}
	\label{fig:tag}
\end{figure}

\subsection{Face-to-face traces}
Both studies involved the use of wearable RFID tags for the collection of face-to-face interactions and location information. Our aim was to use a technology that was not obtrusive, and would not affect significantly the normal social patterns of the participants. 
The measurements were captured by active RFID badges~\cite{Cattuto10:Dynamics}, worn on the body as shown in Figure~\ref{fig:tag}. The badges are lightweight radio transceivers, programmed to transmit a beacon periodically (every 1 second), and to listen continuously for beacons from other badges nearby. The badges are configured to transmit low signal strength beacons that were experimentally evaluated to have a range of 1.5m - 2m with clear line-of-sight. When worn by participants, the beacons are shielded by the body, meaning that successful communication can occur only when another badge is facing that of the participant.  This way the tags can assess continued face-to-face proximity between users. 

We assume continued face-to-face proximity to be a good proxy for a social interaction between users. Specifically, we consider an indication of social interaction the presence of two individuals facing each other at a distance of no more than 2 meters, for a duration of more than 30 seconds. Defining the distance threshold for such matching to be 2m (the configured range of the radio transmission) makes the likelihood of false positives in the dataset negligible. Reducing the number of false negatives (face-to-face proximity not detected by the tags) can be controlled by using time windows within which detected beacons can be considered as indicators of proximity for that duration~\cite{Panisson11:Dynamics}. In this work we use a 2-minute time window, though we verified that 5-minute and 10-minute windows did not significantly alter our findings. We consider as contacts only traces where at least 2 beacons are received 30 seconds apart, thus avoiding counting very short contacts, such as when two people pass one another in the corridor without stopping.

This technology allowed us to capture timestamped contacts of pairs of people. The short communication range of the RF tag (2m) meant that the dataset would not include contacts between larger groups of people interacting over a larger space (e.g. in a meeting). We compensate for this by applying the transitivity property over the original dataset: if participant $P1$ is in contact with participant $P2$, and at the same time participant $P2$ is in contact with participant $P3$, then $P1$ and $P3$ can be considered to have been part of the same group interaction.

\subsection{Location traces}
In addition to face-to-face interactions, the data collection deployment involved the capture of location information. A number of RF tags were deployed around the target buildings. These tags were configured to transmit RF beacons at larger signal strength, achieving a range of around 3m - 5m. Location tags were deployed in every participant's office, in meeting rooms, laboratory spaces, and in communal areas such as cafes, kitchens, and common rooms. These tags allowed us to capture a second dataset that can be used to approximate the locations where social interactions were taking place. 

We establish the approximate location where an interaction takes place by considering the traces of static tags received by all participants in a meeting. We applied a simple voting scheme over the number of static tag beacons received by all interacting users within the specified time window. Using the ID of the static tag with the highest number of beacons received, we then assigned the type of this location (office, meeting room, cafeteria, etc) to that interaction. In the vast majority of our traces the most probable location was clearly distinguishable, with a significantly higher number of beacons received from a single location. In the remaining cases we observed that the potential alternative locations were of the same type as the top one (mostly different nearby offices). Therefore, using the static tag with the highest received beacons allowed us to identify the type of the location where a meeting took place, even though in some cases we could not pinpoint accurately the exact location.

\subsection{Deployments}

The aim of the deployment was to capture a snapshot of the company employees' social behavior. The whole company includes approximately 230 employees.  We sampled our participants from three different research groups, ensuring proportional representation from all layers of the organizational hierarchy. The recruitment process involved an open invitation to all the members of the research groups, followed by a face-to-face consultation where each potential participant had a chance to discuss the details of the study. We recruited 40 participants for both studies; 3 people declined to participate. Each participant signed a consent form, which explained that all published data would be anonymised, including a clause stating that they could withdraw from the study at any point, or have any portion of their data removed from the dataset. We received no requests for data removal. During the study, we did not find the need to enforce the wearing of badges; the fact that people were wearing the badge around the lab acted as a reminder to those who might have forgotten to wear the badge.
In the post-deployment analysis, we cleaned the dataset, removing participants who were not present in both studies, resulting in traces from 24 employees. It is the data from these 24 employees present in both studies that we use in the analysis presented in this paper, in order to make comparisons between their interactions in the two workplace buildings.

The two deployments took place at appropriate times before and after the company's move. As is the nature of the research lab, people may have trips from time to time (work-related or vacations). The times of the two deployments were chosen to be such that not many people were away.
The first data collection deployment was performed in November 2012, where participants were tracked for 2 working weeks. 59 location tags were deployed across 3 floors of the target building. The deployment captured 1669 unique face-to-face contact occasions. The company moved to the new building in January 2013, and the second study was conducted in June 2013, allowing enough time for the participants to settle in the new environment. The second study again collected traces for approximately 2 weeks. 84 location tags were deployed covering 3 floors of the new building. The deployment captured 2693 unique face-to-face contact occasions. Finally, both deployments were complemented with a number of RFID readers that were necessary for collecting data from the wearable RFID tags. Readers were deployed with approximately 30\% overlap in coverage (experimentally measured), in order to minimize the number of lost packets from the wearable tags. 

\begin{figure}
	\centering
	\includegraphics[width=.7\columnwidth]{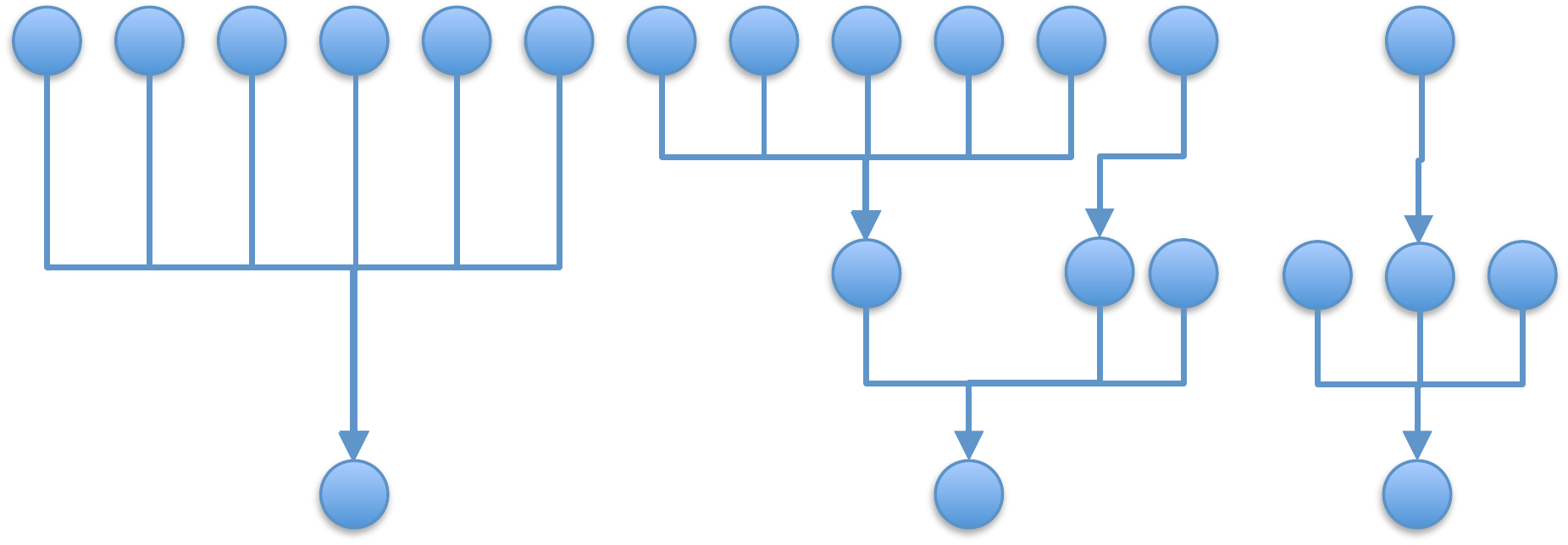}
	\caption{The structure of the subset of the organizational management hierarchy made up of the individuals participating in the study. Circles represent the participants, and arrows indicate the `is managed by' relationship.}
	\label{fig:hierarchy}
\end{figure}

\section{Analysis}

The two deployments resulted in two datasets, consisting of timestamped tuples of user pairs, the type of location where they interacted, and the duration of the interaction. In our analysis we deliberately avoided collecting user-reported data through questionnaires or interviews. Our aim was to attempt to address the research questions without relying on self-reported information. The only additional information used in this analysis was the formal organizational structure of the company, and the layouts of the two buildings.

\subsection{Research Question 1:}
\emph{To what extent is the vertical structure of the formal organizational hierarchy reflected by face-to-face interaction patterns?}

We first examine any correlation between individuals' positions in the vertical levels of the formal organization chart and the number of others they meet.  We make use of the subset of the organization's official management hierarchy that contains the individuals participating in the study, the structure of which is shown in Figure \ref{fig:hierarchy}. We quantify how much this hierarchy manifests in measured interaction patterns by computing for each two-week measurement period (in the old building, and in the new building) the correlation between the degree of individuals in the graph representing the management hierarchy, and their degree in the contact graph constructed by representing each participant by a node, and placing an edge between two nodes when the corresponding participants recorded a face-to-face contact during the study.

Note that in the management hierarchy graph, the degree of a node representing an individual is the number of people in the study who report to that individual or to whom that individual reports. We would expect that in general, if the management hierarchy is strongly manifested in interaction patterns, there would be significant positive correlation between the degree centralities in the management graph and in the contact graphs; this would imply that the more people an individual has reporting to them, the more people they meet face-to-face. Alternatively, if there is \emph{not} significant correlation between individuals' degree centralities in the two graphs, it would imply that the vertical structure of the organization is not so influential in dictating who meets whom (although horizontal structure, as concerned by our next question, may still have an effect).

\subsection{Research Question 2:}
\emph{Is there more face-to-face communication between the different subgroups, as defined by the management hierarchy, in the new building than in the old building?}

We again use the management structure shown in Figure \ref{fig:hierarchy} to define subgroups, and consider three such groups, one corresponding to each of the three components present in the graph. We first quantify inter-group contact by measuring the proportion of contact pairs that are intra- and inter-group: we expect that there will be a higher proportion of contact pairs inter-group in the new building, given the emphasis in the building's design on shared spaces to facilitate inter-group contact.

We further analyze communities in the contact graphs. In network theory, a community is defined to be a group of nodes in the network with particularly many or dense connections between them, and fewer or looser connections to the nodes not in that group. In terms of the flow of ideas and information, a less modular structure or one with more connections between communities would be advantageous~\cite{Fortunato10:Community}.

To measure these properties, we first find $k$-clique communities in the network, defined to be the union of all cliques of size $k$ in the graph that can be reached through adjacent (sharing $k - 1$ nodes) $k$-cliques~\cite{Derenyi05:Clique}, and compute the proportion of edges in the contact graph that exist within and between communities, for varying values of $k$. We would expect that, if the new building space promotes inter-group interaction as the architects intended, the proportion of inter-community edges would be higher for the contact graph from the new building than that for the graph from the old building.

We also run the Louvain algorithm for community detection~\cite{Blondel08:Fast} on the contact graph, which partitions the network into communities and enables us to compute the modularity $Q$ of this partition, a measure of the strength of the community structure of the network~\cite{Newman06:Modularity}. The value of $Q$ may range between -1 and 1, with a value of 0.3 or more considered high. We expect that in the new building there would be a less strong community structure to the contact graph due to increased mixing between different groups of people, and therefore the value of $Q$ would be lower than that seen in the old building.

\begin{figure*}
        \subfigure[\% of contact pairs occurring inter-group]{
		\includegraphics[width=\columnwidth]{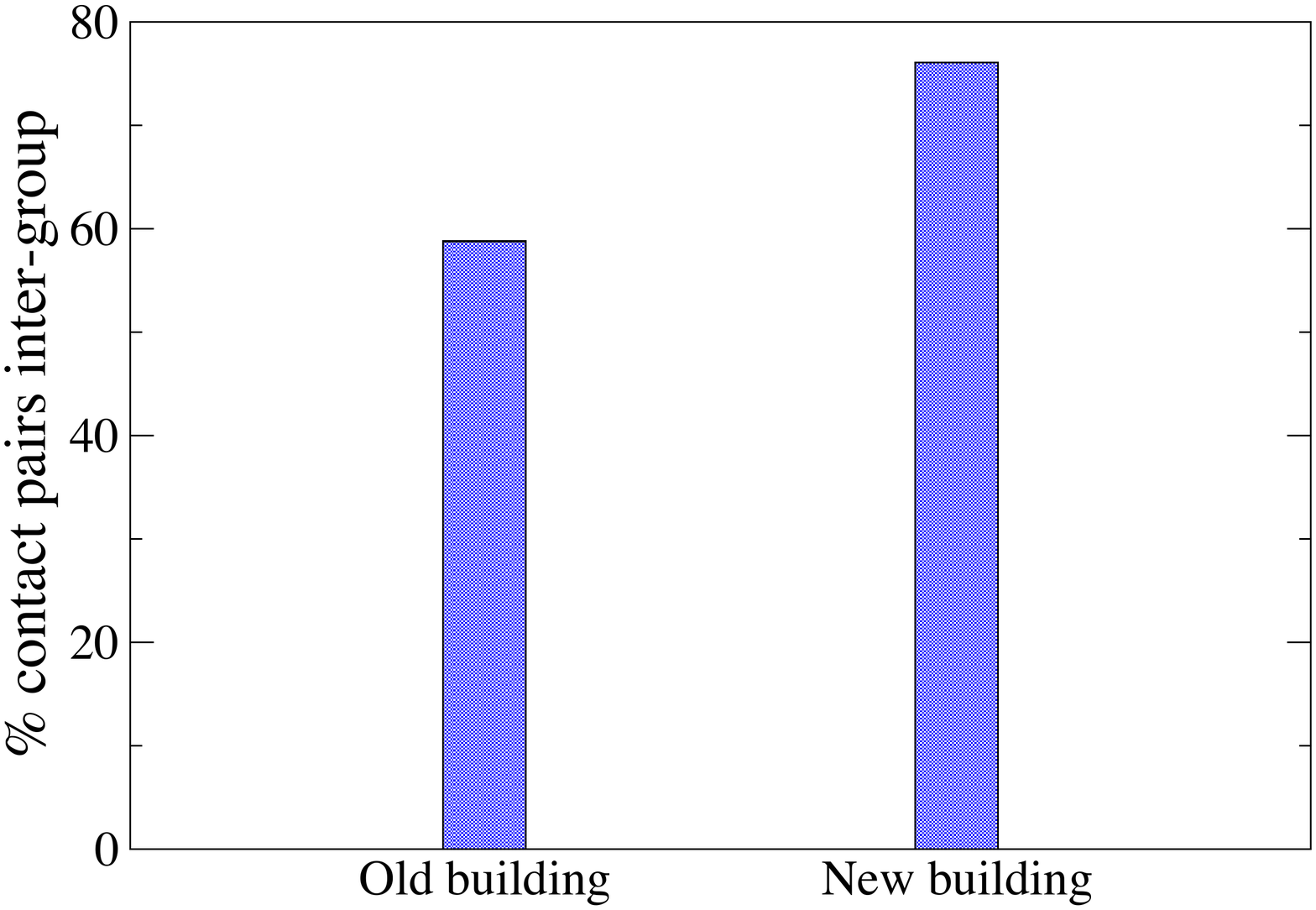}
		\label{fig:percentages}
        }
        \subfigure[Old building netgraph]{
		\includegraphics[width=0.75\columnwidth]{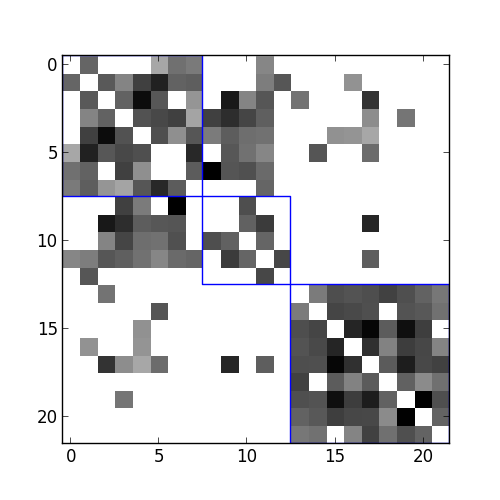}
		\label{fig:old-netgraph-all}
       }
        \subfigure[New building netgraph]{
		\includegraphics[width=0.75\columnwidth]{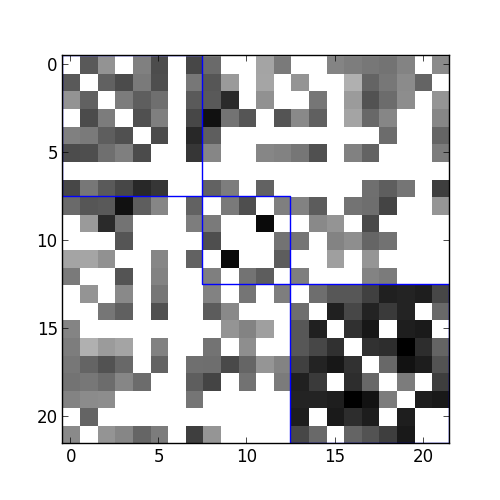}
		\label{fig:new-netgraph-all}
       }
        \caption{Inter- and intra-group contact pairs in the old and the new buildings.
        In the netgraphs, each row and column corresponds to an individual, and the ordering of individuals is by group so that adjacent rows (columns) represent colleagues in the same group. Group boundaries are shown by the blue lines. A dark square indicates that contact was recorded between the individuals concerned, and a white square indicates no recorded contact. A more `blocky' structure suggests less inter-group mixing. The total duration of recorded contact is reflected by how dark a shaded square is, with the durations normalized by the total time that the individuals concerned were recorded as being in the building and also according to the maximum recorded duration. The increase in inter-group contacts in the new building is clear.}
        \label{fig:netgraphs-all}
\end{figure*}

\subsection{Research Question 3:}
\emph{Does mixing between people in different subgroups takes place in food and drink areas?}

It is well-known that areas in which people can gather together to eat and drink are often important hubs for the kind of social contact and chance meetings that are so beneficial for the exchanges of ideas and information~\cite{Allen06:Organization,Isaacs96:Piazza}. Therefore it is unsurprising that the architects of the new building in our study envisaged that the cafeteria would promote interaction between people in different groups who might not normally encounter one another.

We test quantitatively the importance of food and drink spaces for inter-group interactions by computing the number of contacts taking place in different kinds of spaces (e.g., offices, meeting rooms, kitchens) within the two buildings over the course of the working day. We would expect that in the new building, we would see the impact of the cafeteria manifest as an increase in the number of contacts occurring over the lunch hour (12-1pm). We further investigate the importance of lunchtime for inter-group contact by comparing the proportion of contact pairs that are between individuals from different groups in all of the data for each building, and in the same data but with the lunch hour removed. We would expect that in both cases, the proportion of communicating inter-group pairs goes down when lunchtime is excluded from the analysis.

We then specifically investigate the proportion of inter- and intra-group contacts that occur in different kinds of spaces in the old building and in the new building. We would expect that in the new building, a greater proportion of inter-group contacts would occur in kitchen areas (which include the cafeteria) than in the old building, and also that these proportions would reflect the inclusion of more meeting rooms in the new building to encourage colleagues to venture away from their own offices in order to hold work-related meetings.

\section{Results}
\subsection{Research Question 1:}
\emph{To what extent is the vertical structure of the formal organizational hierarchy reflected by face-to-face interaction patterns?}

Table \ref{tab:degree} shows the value of Pearson's correlation coefficient $r$, and the corresponding $p$-values, for the degree of individuals in the formal organizational management graph (as shown in Figure \ref{fig:hierarchy}) vs.~degree in the contact graph, for both the old and the new buildings. In the old building, $r=-0.17$, showing weak negative correlation, but this is not statistically significant, with $p=0.50$. In the new building, there is virtually no correlation at all between the two degree values, with $r=-0.04$ and $p=0.89$.  We see that \emph{the apparent irrelevance of the formal management structure for dictating how many others people interact with was preserved by the change in building spaces}.

\begin{table}
\begin{centering} 
\begin{tabular}{|l|c|}
\hline
& $r$ \\ \hline
\textbf{Old building} & -0.17 (0.50) \\ \hline
\textbf{New building} & -0.04  (0.89) \\ \hline
\end{tabular}
\caption{Pearson's correlation coefficient $r$ (and $p$-value) for degree in the management graph vs.~degree in the contact graph. There is not significant correlation between the vertical organizational structure and meeting patterns.}
\label{tab:degree}
 \end{centering}
\end{table}
The lack of significant correlation between individuals' degree in the management graph and in the contact graphs shows that individuals' vertical positions in the formal organizational hierarchy has little effect on the number of others they come into contact with. We note that this may be because, as demonstrated in a recent study of the relationship between physical workspace and communication by Steelcase~\cite{Steelcase13:How}, in the UK it is common for managers to ``[invite] interaction among employees at all levels", reducing the impact of the vertical levels of formal organizational structure on face-to-face encounters. The fact that this effect persisted between the old and the new workplace buildings may mean that \emph{communication for coordination}, as Thomas Allen terms formal meetings between direct colleagues to organize work, was not strongly affected by the physical building space. 

However, informal or unplanned meetings between people separated horizontally into different subgroups of the formal organizational structure may be more affected by space~\cite{Henn13:Serendipitous}, which would present problems for \emph{communication for inspiration}. In investigating our next research question, we proceed to test whether this is the case.

\begin{figure}
    \centering
    \includegraphics[width=\columnwidth]{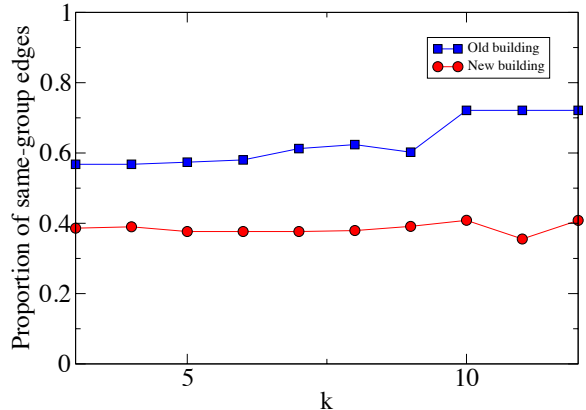}
    \caption{Proportion of intra-community edges that are between individuals in the same group, for varying values of $k$ used in the $k$-clique algorithm for community detection. The proportion of same-group intra-community edges is lower in the new building than in the old building.}
    \label{fig:kcliques}
\end{figure}

\subsection{Research Question 2:}
\emph{Is there more face-to-face communication between the different subgroups, as defined by the management hierarchy, in the new building than in the old building?}

Figure \ref{fig:percentages} shows the percentage of contact pairs that involved people in different groups, where groups are defined to be the three distinct subtrees of the management structure shown in Figure \ref{fig:hierarchy}. In the new building, 76.1\% of pairs were cross-group, increased from 58.8\% in the old building. This suggests that \emph{the aim of the architects to design the building to promote mixing between people who might not encounter one another otherwise may have been successful}, and that there might be more opportunities for \emph{communication for inspiration} in the new building.

Figure \ref{fig:netgraphs-all} shows a visual representation of the extent of this effect, in the form of netgraphs~\cite{Allen06:Organization}. Each row and column corresponds to an individual, and the ordering of individuals is by group so that adjacent rows (columns) represent colleagues in the same group. Group boundaries are shown by the blue lines. A grey square indicates that contact was recorded between the individuals concerned, and a white square indicates no recorded contact. A more `blocky' structure suggests a lower level of inter-group mixing.

The netgraphs make clear the extent to which \emph{more inter-group mixing is encouraged by the design of the new building}, with many more dark squares outside the blue lines indicating contact between individuals in different formal subgroups.

\begin{figure}
    \centering
    \includegraphics[width=0.6\columnwidth]{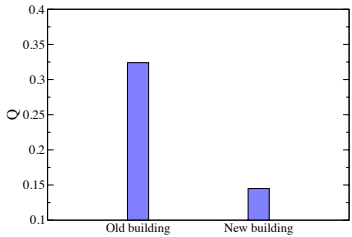}
    \caption{Modularity $Q$ of the best partition of the contact graph into communities, as found by the Louvain algorithm. A larger value indicates a stronger community structure, with values of around 0.3 or more being considered high. The community structure of the contact graph in the new building is less strong, suggesting more encounters between individuals from different communities.}
    \label{fig:modularity}
\end{figure}

\begin{figure}
    \subfigure[Inter-group contacts in old building]{
        \includegraphics[width=0.6\columnwidth, height=0.46\columnwidth]{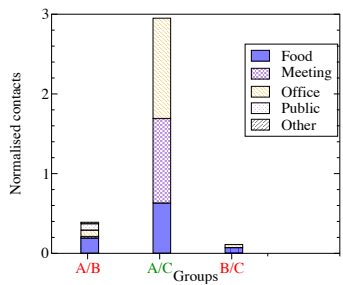}
		\label{fig:inter-group-old-building}
}
    \subfigure[Inter-group contacts in new building]{
        \includegraphics[width=0.6\columnwidth, height=0.46\columnwidth]{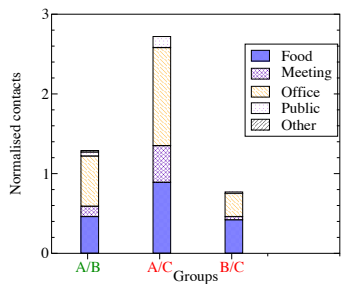}
		\label{fig:inter-group-old-building}
}
    \caption{The labels on the horizontal axis show the pairs of groups, marked as green for groups located on the same floor, and red for those on separate floors.}
    \label{fig:inter-group-floors}
\end{figure}

\subsubsection{Community analysis}
Figure \ref{fig:kcliques} shows the results of $k$-clique analysis~\cite{Derenyi05:Clique} of the contact graph. Specifically, we plot, for varying values of $k$, the proportion of intra-community edges in the contact graph that are between individuals in the same subgroup.

We see that the proportion of intra-community edges connecting those in the same group is lower in the new building than in the old building, for all of the values of $k$. This implies that the community structure of the contact graph is less constrained by the formal group structure and therefore that \emph{the new building space may indeed encourage more opportunities for mixing between individuals in different groups}.

We further confirm this result by checking the modularity $Q$ of the best partition of the contact graphs found by running the Louvain community detection algorithm~\cite{Blondel08:Fast}. A larger value indicates a stronger community structure, with values of around 0.3 or more being considered high. Figure \ref{fig:modularity} shows the value of $Q$ for the contact graphs in the old and the new buildings; we can see that $Q$ is lower for the contact graph in the new building, which suggests again that \emph{the community structure is less modular and that there are more contacts that are outside usual meeting groups}.

\subsection{Research Question 3:}
\emph{Does mixing between people in different subgroups takes place in food and drink areas?}

\begin{table}[b]
\begin{centering} 
\begin{tabular}{|l||c|c|}
\hline
& \textbf{Old building} & \textbf{New building} \\ \hline
\textbf{Group A} & Floor 1 & Floor 3\\ \hline
\textbf{Group B} & Floor 2 & Floor 3\\ \hline
\textbf{Group C} & Floor 1 & Floor 1\\ \hline
\end{tabular}
\caption{Distribution of groups on different floors in the two buildings.}
\label{tab:floor-allocations}
\end{centering}
\end{table}

Before examining the impact that different location types may have on inter-group interactions, we assess the extent to which the distribution of  offices across floors may affect interactions. Previous work has indicated that splitting employees across floors may have a significant impact in social interactions, mostly in traditional building designs~\cite{Sailer11:Social}. Of the individuals involved in this analysis, each was on the same floor as the rest of their group, but some groups shared a floor and others were on different floors. (Table \ref{tab:floor-allocations}).

\begin{figure*}
        \centering
        \subfigure[\% inter-group contacts]{
		\includegraphics[width=\columnwidth]{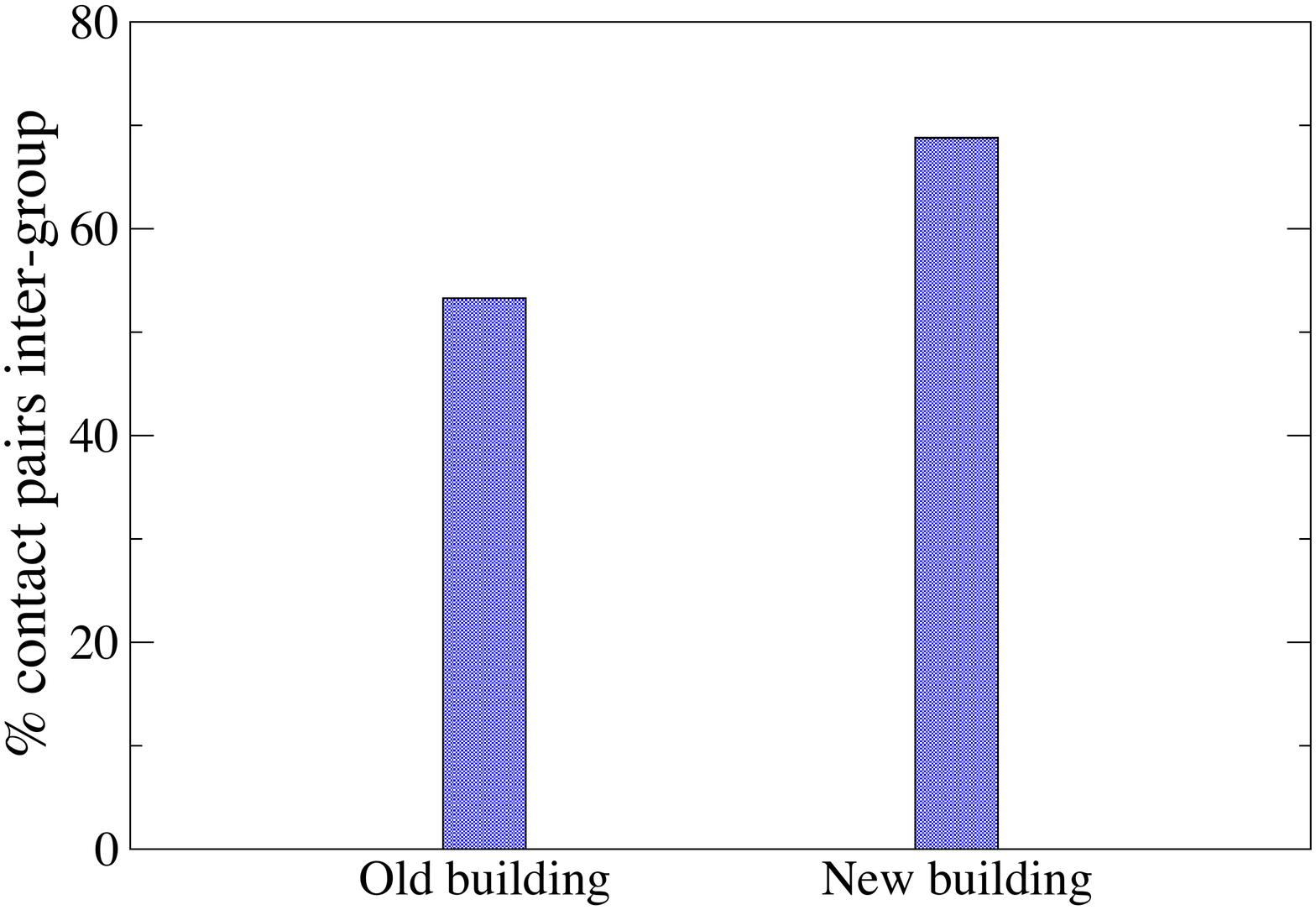}
		\label{fig:inter-group-no-lunchtime}
        }
        \subfigure[Old building]{
		\includegraphics[width=0.75\columnwidth]{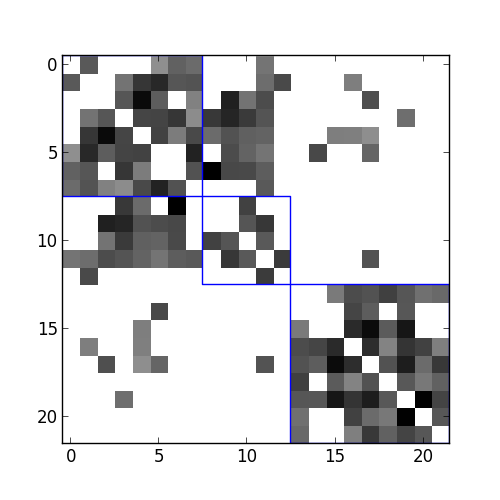}
		\label{fig:old-netgraph-no-lunchtime}
        }
        \subfigure[New building]{
		\includegraphics[width=0.75\columnwidth]{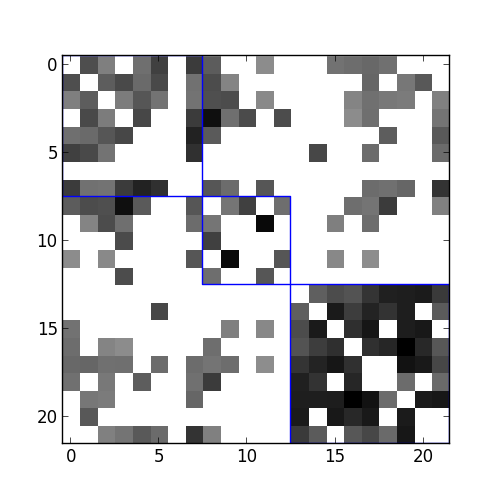}
		\label{fig:new-netgraph-no-lunchtime}
        }
        \caption{Inter- and intra-group contact pairs in the old and the new buildings, with lunchtime removed. The total duration of recorded contact is reflected by how dark a shaded square is, with the durations normalized by the total time that the individuals concerned were recorded as being in the building and also according to the maximum recorded duration. The proportions of inter-group contacts are lower than those shown in Figure \ref{fig:percentages}, which demonstrates the importance of lunchtime for social contact in both buildings.}
        \label{fig:netgraphs-no-lunchtime}
\end{figure*}

In order to assess how this distribution may affect interactions we calculated the number of inter-group interactions for each pair of groups in both buildings. The values were normalized by the product of the sizes of the groups involved, to account for the number of possible pairs: $N_{AB}= \frac{C_{AB}}{|A| \cdot |B|}$ \\
where $N_{AB}$ is the normalized inter-group contacts for groups $A$ and $B$, $C_{AB}$ is the absolute number of inter-group contacts, and $|A|$ and $|B|$ are the sizes of the two groups. Figure~\ref{fig:inter-group-floors} shows the normalized number of inter-group contacts in the two buildings. Note that group pairs A-C in the old building, and A-B in the new building are located on the same floor.

\begin{figure}
        \centering
        \subfigure[Old building]{		\includegraphics[width=0.75\columnwidth, height=0.5\columnwidth]{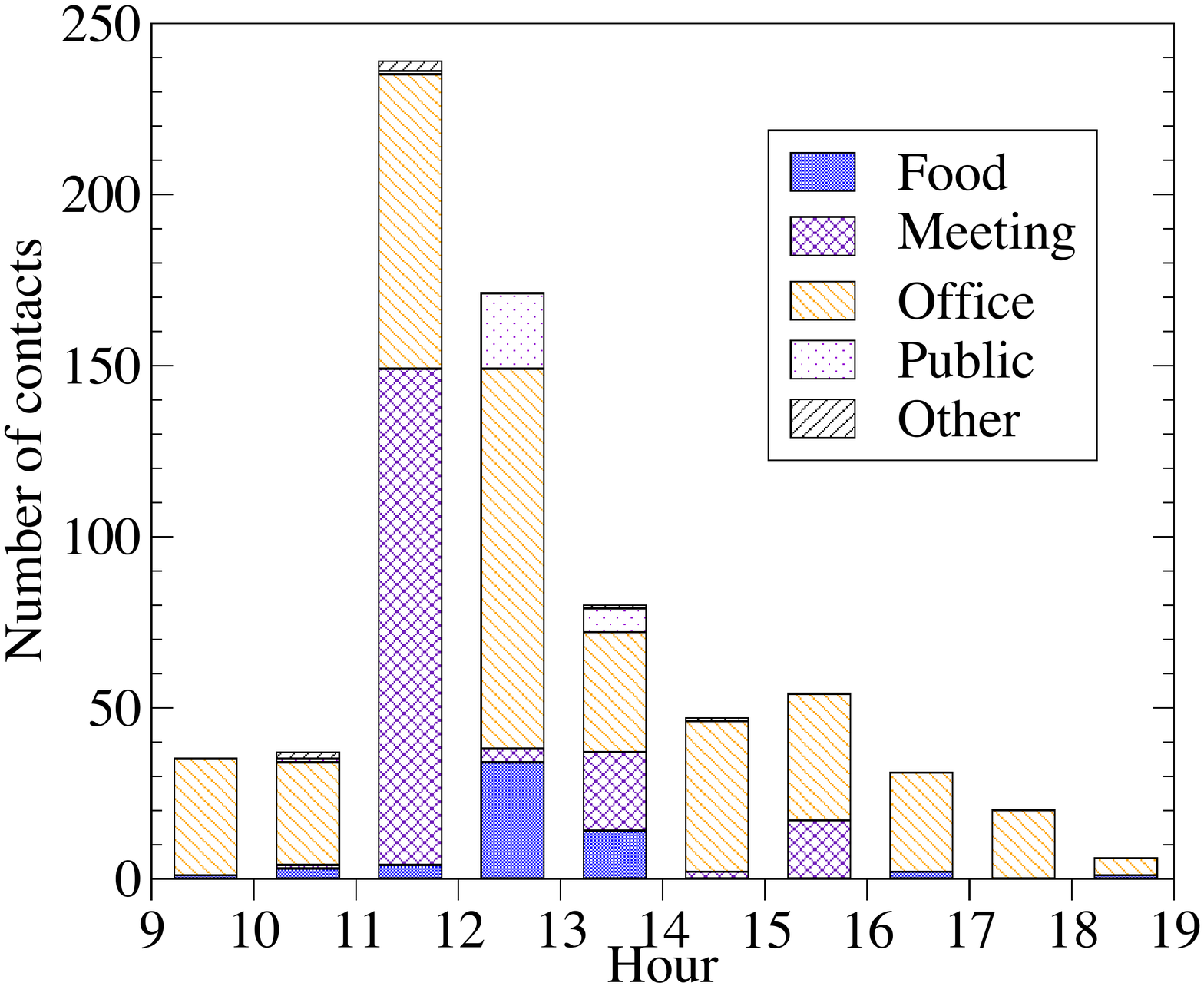}
		\label{fig:old-times}
       }
        \subfigure[New building]{
		\includegraphics[width=0.75\columnwidth, height=0.5\columnwidth]{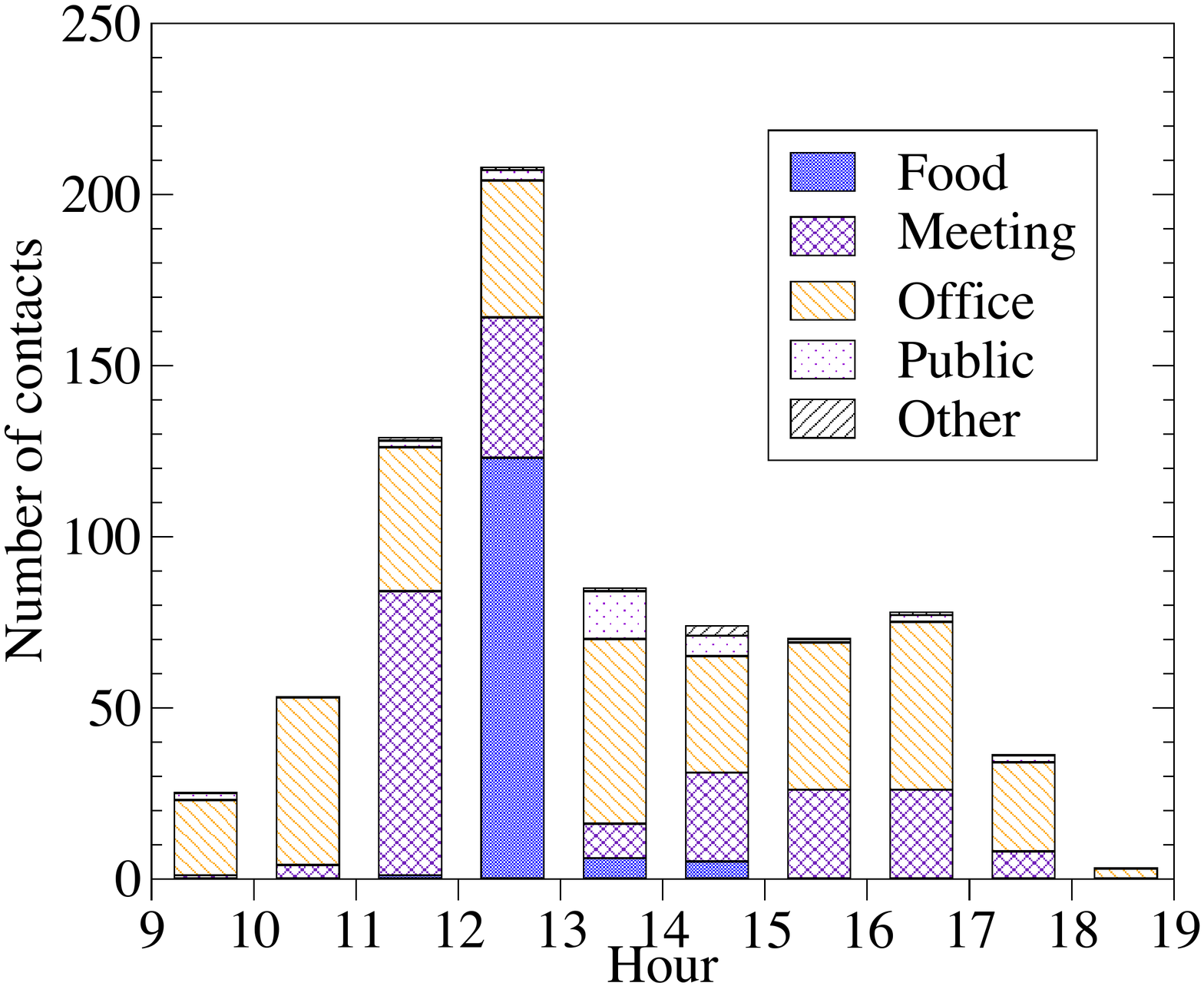}
		\label{fig:new-times}
       }
        \caption{Number of contacts in different kinds of spaces, by hour of the day. The impact of the cafeteria in the new building is clearly visible, with a large increase in contacts in Food spaces between the hours of 12-1pm.}
        \label{fig:times-all}
\end{figure}

The graphs show that in the old building, floor allocations are a strong factor in inter-group interactions. Indeed, the vast majority of inter-group interactions is found between groups A and C, with offices on the same floor. In the new building, we can see an increase in inter-group interactions for all the pairs. Interestingly, groups A and C maintain the same level of interaction although they are now located on different floors. Furthermore, groups B and C show a 7-fold increase compared to the old building, although they are located on separate floors. This increase is higher than the 3.5-fold increase in interactions between groups A and B, which are located on the same floor. These results suggest that in the new building the distribution of offices across floors is not the dominant factor determining the interactions between groups. The graphs also suggest that interactions in food areas may be more important in the new building.

We now examine the impact of lunchtime on inter-group interactions, by comparing the proportions of contact pairs comprising people in different groups previously shown (Figure \ref{fig:percentages}) with the same excluding lunchtime (12-2pm) from the analysis. Figure \ref{fig:inter-group-no-lunchtime} shows the results; in both the old and the new building, the proportion of inter-group contact pairs decreases when lunchtime is excluded, demonstrating that \emph{lunchtime is, as expected, important for contact between people in different groups}. The netgraphs constructed from contacts excluding lunchtime (Figures \ref{fig:old-netgraph-no-lunchtime} and \ref{fig:new-netgraph-no-lunchtime}) further confirm this result.

We next investigate the kinds of spaces where contacts take place, by plotting the numbers of contacts detected between participants in different kinds of spaces (food and drink areas, meeting rooms, offices, and public areas) during different hours of the day (Figure \ref{fig:times-all}). The impact of the cafeteria in the new building is clearly visible as a sharp increase in the number of contacts taking place in Food spaces during the hours of 12-1pm. In conjunction with the previous finding, that this time of day is important for inter-group contacts, this analysis does suggest that \emph{food and drink areas are important for mixing between people in different subgroups}.

Finally, we examine directly the proportion of inter-group contacts that occur in each kind of space.
Table \ref{tab:inter-group-locations} shows the results from each building. We can see that in the new building, a greater proportion of inter-group contacts take place in food areas than before, which suggests again that \emph{the intention of the architects for the cafeteria to function as somewhere where people in different groups have the opportunity to meet one another has been realized}.

Furthermore, the proportions of inter-group contacts that take place in Food spaces far exceed the proportions of intra-group interactions that take place in Food spaces in both buildings: in the old building, while only 3.7\% of intra-group interactions take place in Food spaces, 28.0\% of inter-group interactions happen there. In the new building, 9.4\% of intra-group contacts occur in Food spaces, compared with 37.4\% of inter-group interactions. This demonstrates the importance of food and drink areas for inter-group encounters.\\

The table also shows the proportions of \emph{intra-}group contacts that take place in each of the different kind of spaces. We can see that the majority of these contacts occur in meeting rooms and offices, which is expected given that such contacts are likely to be those comprising what Thomas Allen refers to as \emph{communication for coordination} and \emph{communication for information}. Interestingly, we also note that more of these contacts take place in meeting rooms in the new building than in the old building, which suggests \emph{that the architects' aim of encouraging more meetings in shared spaces, away from individuals' own offices, was indeed met in the realization of the design}.\\

\begin{table}
\begin{centering} 
\begin{tabular}{|l||c|c|}
\hline
&  \textbf{Old building (\%)} & \textbf{New building (\%)} \\ \hline
\textbf{Food} & 28.0 (3.7) & 37.4 (9.4) \\ \hline
\textbf{Meeting} & 28.8 (24.4) & 12.3 (34.0) \\ \hline
\textbf{Office} & 38.7 (65.5) & 45.5 (51.5) \\ \hline
\textbf{Public} & 3.8 (5.1) & 4.3 (3.7) \\ \hline
\end{tabular}
\caption{\% of inter-group (and intra-group, in parentheses) meetings that take place in each kind of space, in the old building and in the new building. Inter-group contacts far exceed intra-group interactions in Food spaces.}
\label{tab:inter-group-locations}
 \end{centering}
\end{table}

\section{Discussion}
This work fills a gap in the body of studies involving the interaction between physical space and formal organizational structure to influence face-to-face encounters between individuals in the workplace. Specifically, we have exploited the advantages of modern ubiquitous sensing technology over methods such as manual observation and self-reports to measure face-to-face encounters, in a direct comparison of the communication behavior of the same employees from the same organization in two different physical workplace buildings. 

It is usually difficult to obtain suitable data for studies such as this one due to the infeasibility of simply moving an organization from one building to another for the purposes of an experiment, owing to the high time, effort, and financial costs, and the fact that studies comparing organizations in different kinds of buildings cannot account for all the possible organization-specific variables that might affect the validity of comparisons. Furthermore, while it is possible to study the effect of spatial configuration within a single organization at lower cost, by simply changing office layout, for example, a fundamental aspect of our study is the nature of the spaces provided by the building (e.g.~number and location of meeting rooms, food and drink areas, etc.), which cannot be changed without altering the building itself.

\subsection{Theoretical implications}
This study provides a rare example of a direct comparison of two different workplace buildings and the impact of the space on the communication of and potential for interaction between the same employees, in conjunction with the formal organizational structure. We provide evidence building on the body of existing work on this subject, supporting the idea that communal spaces could be important to provide opportunities for \emph{communication for inspiration} between employees who may not work together and are in different organizational subunits, and that there is demonstrably more potential for encounters in these spaces between those who may not otherwise meet. We also see that office and meeting room spaces are most likely still important for \emph{communication for coordination} and \emph{communication for information} between members of the same team, showing directly the value of both kinds of spaces to allow all of the forms of communication important for a thriving innovative organization. 

Furthermore, we have shown that the data suggest that the aims of the architects to encourage more use of shared spaces -- both by members of the same organizational subunits, exemplified by the provision of more meeting rooms in the new building, and by members of different teams, as in the case of the cafeteria -- were met, which also provides evidence that such architectural considerations can be of value.

\subsection{Practical implications}
We have demonstrated the use of ubiquitous sensing technology to monitor changes in face-to-face communication patterns between employees influenced by the spaces provided by their workplace building. It is easy to see that such information could be used fruitfully to provide feedback to the employees themselves on their patterns of communication, or indeed to their managers, to encourage people to take advantage of the possibility of encounters with those they might not normally meet or opportunities to interact informally with their teammates. 

To give an example of such a practical application, it would be possible to create maps of workplace floor plans reflecting the extent to which each area hosts conversations between those at different positions in the formal organizational structure, and therefore which kinds of communication (for coordination, information, or inspiration) are most likely to take place in which kinds of space. Similarly, it would also be possible to monitor using the same technology the kinds of interaction in which an individual engages over the working day, and to make this information available to the employees, to encourage them to balance the three kinds of communication to meet their working needs. For example, one could create graphs showing how a research group is connecting within their own team and also how well-connected they are with other teams, both of which have been shown to be important in the workplace~\cite{Pentland12:New}. It would then be possible to measure the impact of this feedback on organizational productivity, adding to branches of research investigating how ubiquitous computing can be used to nudge people to change their behavior~\cite{Rogers10:Ambient}.

Of course, the ability to track social interactions in the workplace and the potential display of data from this monitoring raises further questions about privacy, how happy employees would tend to be about being tracked in this way with the potential for others to see their data, and whether this in itself would cause behavior to change. All of these issues could be ground for further research before this type of technology became widely deployed in workplaces.

\subsection{Limitations}
The requirements for recording a contact using the active RFID badges are fairly stringent, and this may mean that while it is possible to mitigate the problem of false negatives (failing to record an encounter when one takes place) as outlined above, we can still underestimate the levels of contact occurring. However, we note that this issue is consistent across the two measurement periods in the two buildings, since we set up the experiment in the same way using the same technology, and so the comparisons we draw are still valid, despite the fact that absolute numbers of contacts should not be taken as completely accurate. Similarly, we could not record contacts taking place outside the building, and indeed did not aim to; we aim here to examine not the absolute levels of contact between employees, but how the spaces of the workplace building are used for such interactions.

One should also bear in mind that this is just one sample of one organization, and should not be taken as representative. Different organizations might be affected differently under the same conditions; many more such studies would be needed in order to draw more general conclusions. Furthermore, since this specific organization is a research lab, it implies a particular way of interaction both horizontally and vertically in the organizational structure that might be different in other types of organization, for example, those that are more commercial rather than research-oriented. Again, our results should not be assumed to be representative.

Further, while we have considered communication between employees according to whether it is likely to be for co-ordination, for information, or for inspiration, clearly it is not possible to determine the actual content of conversations that took place, due to the measurement technology used. We emphasize that the conclusions we draw reflect the likely potential for communication of these kinds according to the kinds of spaces where interactions take place and the positions of the people involved in the organization, rather than being intended to make direct claims about topics of conversation.

Finally, this study concerns only the short-term impact of the different physical workplace environment on face-to-face communication. Other studies have shown that such face-to-face interaction between employees can have important effects on productivity and innovation, but these are phenomena that require evaluation over a longer period. In the first instance we have dealt only with communication patterns measured over the short term, and not their consequences, but we would be able in the future to examine these effects in this case, perhaps by considering in the case of the research institution concerned metrics involving collaborations and publications.

\section{Conclusions}
We have studied a unique dataset of contact traces captured using active RFID badges to sense face-to-face encounters between employees of a research institution in the UK, during two weeks in one building and during another two weeks after the organization had moved to a different building. We have thus been able to analyze how the formal organizational structure and the physical space of the working environment combine to affect communication between employees. We have provided empirical evidence for the importance of informal spaces in providing opportunities for \emph{communication for inspiration} between employees working in different subgroups, as well as demonstrating that \emph{communication for coordination} and \emph{communication for information} seem likely to happen in office and meeting spaces between members of the same teams. 

This is, to the best of our knowledge, the first time that such a study using purely automatic data collection methods has been conducted to perform direct comparisons between the behavior of the same group of employees in the same organization in two different workplace buildings, thus allowing us to study directly the effect of the physical building space without the interference of organization-specific variables that may be involved when making comparisons across different institutions.

\section{Acknowledgements}
We thank all of the participants who took part in the study. We are also grateful to Kerstin Sailer for useful feedback and discussion. Chlo\"e Brown is a recipient of the Google Europe Fellowship in Mobile Computing, and this research is supported in part by this Google Fellowship.

\end{document}